\documentclass[prb,preprint,showpacs,showkeys]{revtex4}
\usepackage{graphicx}
\usepackage{amsmath}
\usepackage{amssymb}
\usepackage{bm}

\begin{document}


\title{Nonlinear Seebeck and Peltier Effects in Quantum Point Contacts}

\author {M. Ali \c{C}\.IP\.ILO\u{G}LU}
 \email {ali.cipiloglu@bayar.edu.tr}
 \affiliation {Physics Department, Celal Bayar University, Manisa, TURKEY}

\author {Sadi TURGUT}
 \email {sturgut@metu.edu.tr}
\affiliation{Department of Physics, Middle East Technical
University, ANKARA, TURKEY}

\author{Mehmet TOMAK}

\affiliation {Department of Physics, Middle East Technical
University, ANKARA, TURKEY}

\date{\today}


\begin{abstract}
The charge and entropy currents across a quantum point contact is expanded as a
series in powers of the applied bias voltage and the temperature difference.
After that, the expansions of the Seebeck voltage in temperature difference and
the Peltier heat in current are obtained. With a suitable choice of the average
temperature and chemical potential, the lowest order nonlinear term in both
cases appear to be of third order. The behavior of the third-order coefficients
in both cases are then investigated for different contact parameters.
\end{abstract}

\maketitle


\section{INTRODUCTION}

Various aspects of the ballistic electron transport across quantum point
contacts are studied extensively in the past. The most striking feature of this
transport is the quantization of conductance\cite{qc1,qc2} at integer multiples
of the conductance quantum $2e^2/h$. This phenomenon is usually treated with
the Landauer-B\"uttiker formalism\cite{landauer,buttiker1} which provides a
transparent explanation for the effect. Electrons in each sub-band
corresponding to the transverse modes in the contact contribute one quantum to
the conductance if the sub-band is sufficiently populated. As the size of the
constriction is changed by varying the negative voltage on split gates, which
are used to define the contact on a two-dimensional electron gas, the
conductance changes in smooth steps from one conductance quantum into the
other. It is observed that the linear Seebeck and Peltier coefficients for
these structures display quantum
oscillations\cite{streda,proetto,houten,molenkamp1,molenkamp2} with peaks
coincident with the conductance steps.

Nonlinear transport in these systems has also been studied extensively both
theoretically\cite{ma:weak,ma:harm,ma:kinetic,qiao,todorov,bogachek1,bogachek2}
and experimentally.\cite{krishnaswamy,dzurak} Since Onsager's reciprocity
relations connecting the Seebeck and Peltier transport coefficients loose its
meaning in this regime, these two effects show distinctively different
behavior. New peaks appear in the differential Peltier coefficient as the
driving voltage is increased,\cite{bogachek1,bogachek2} while the thermopower
does not change much even for very large temperature differences.\cite{dzurak}

A major theoretical difficulty in the nonlinear regime is, due to the small
size of these systems, finite voltage differences create large changes in the
distribution of electrons around the contact. As a result, more involved
calculations are necessary for describing the electron
transport.\cite{buttiker2,christen} However, it is of some interest to analyze the
nonlinear transport properties without taking such changes into account.
The purpose of this article is to
investigate the nonlinearities in
not so commonly studied Seebeck and
Peltier effects, assuming that the contact potential is not changed apart from
the uniform shift caused by the gate voltage.
It is hoped that this will clarify the importance of
the effects mentioned above.
In the following section, the
charge and heat currents are expanded as a series in powers of the potential
and temperature differences. Appropriate expansions for the Seebeck and Peltier
phenomena are obtained and the series coefficients are investigated in sections
2 and 3, respectively. Finally, the results are summarized and discussed.

\section{Theory}

In the following we consider two electron gases connected by a quantum point
contact. The chemical potentials $\mu_L$ and $\mu_R$ and the temperatures
$\theta_L$ and $\theta_R$ of the left ($L$) and right ($R$) reservoirs are the
parameters that define the whole system. The difference between the chemical
potentials, $\Delta\mu=\mu_L-\mu_R$, is equal to $(-e)V$ where $V$ is
interpreted as the electrical potential difference between $L$ and $R$. A
difference in temperatures $\Delta\theta=\theta_L-\theta_R$ as well as a
potential difference cause electron transport which can carry both charge and
heat across the contact. The average currents on the contact are completely
determined by the sum
$$ T(E)=\sum_{n} T_n(E) \quad, $$
where $T_n(E)$ is the transmission probability of an electron with energy $E$
incident from the $n$th mode. The charge and entropy currents from $L$ to $R$
can then be expressed as\cite{form:imry}
\begin{eqnarray}
I &=& 2\frac{(-e)}{h}\int_{-\infty}^\infty dE (f(x_L)-f(x_R)) T(E)\quad,
\label{eq:chcur}\\
I_S &=& 2\frac{k_B}{h}\int_{-\infty}^\infty dE (s(x_L)-s(x_R)) T(E)\quad,
\label{eq:encur}
\end{eqnarray}
where
\begin{eqnarray}
  f(x) &=& \frac{1}{1+e^x}\quad, \\
  s(x) &=& -f(x)\log f(x)-(1-f(x))\log(1-f(x)), \\
  x_{L,R} &=& \frac{E-\mu_{L,R}}{k_B\theta_{L,R}}\quad,
\end{eqnarray}
and the spin degeneracy factor is added for both currents.

For the case of weak nonlinearities, it is useful to expand the currents in
terms of the driving temperature and potential differences $\Delta\theta$ and
$V$. In order to do this the variable of integration is changed from energy $E$
to a dimensionless variable denoted by $\overline{x}$, which is defined as the
arithmetic average of $x_L$ and $x_R$.
$$ \overline{x}=\frac{1}{2}(x_L+x_R)\quad.$$
This leads us to define average temperature and chemical potentials by
\begin{eqnarray}
\overline{x} &=& \frac{E-\overline{\mu}}{k_B\overline{\theta}}\quad,\\
\overline{\theta} &=& \frac{2\theta_L\theta_R}{\theta_L+\theta_R}\quad,
    \label{eq:thetabar}\\
\overline{\mu} &=& \frac{\theta_R\mu_L+\theta_L\mu_R}{\theta_R+\theta_L}\quad.
    \label{eq:mubar}
\end{eqnarray}
Here, $\overline{\theta}$ is the harmonic average of the temperatures of the
two electron gases and $\overline{\mu}$ is an average of chemical potentials
weighted by inverse temperatures. These two quantities will be considered as
the fundamental parameters describing the contact. In other words all of the
transport coefficients are considered as functions of these average quantities.

With these definitions the energy variable can expressed as
$E=\overline{\mu}+\overline{x}k_B\overline{\theta}$ and the difference of the
dimensionless $x$ parameter is
\begin{equation}
  \Delta x = x_L-x_R = -\frac{\Delta\mu+\overline{x}k_B\Delta\theta}{k_B\theta_A}
\end{equation}
where $\theta_A$ is the arithmetic average of the temperatures on both sides of
the contact
$$\theta_A=\frac{1}{2}(\theta_L+\theta_R)\quad.$$
Finally, dimensionless driving forces are defined as
\begin{eqnarray}
  \epsilon&=& \frac{\Delta\theta}{\theta_A}\quad, \\
  \delta &=&  \frac{\Delta\mu}{k_B\theta_A}\quad.
\end{eqnarray}

The obvious advantage of these definitions is the elimination of some terms in
the power series expansion of the integrands in equations (\ref{eq:chcur}) and
(\ref{eq:encur}). We have
\begin{eqnarray}
  \nonumber I &=& 2\frac{(-e)}{h} \sum_{m=0}^\infty \frac{k_B\overline{\theta}}{2^{2m}(2m+1)!}\times\\
    &\times& \int d\overline{x} ~(\delta+\overline{x}\epsilon)^{2m+1} f^{(2m+1)}(\overline{x})
    T(\overline{\mu}+\overline{x}k_B\overline{\theta})\quad,\\
  \nonumber I_S &=& 2\frac{k_B}{h}\sum_{m=0}^\infty \frac{k_B\overline{\theta}}{2^{2m}(2m+1)!}\times\\
    &\times& \int d\overline{x}~ (\delta+\overline{x}\epsilon)^{2m+1} s^{(2m+1)}(\overline{x})
    T(\overline{\mu}+\overline{x}k_B\overline{\theta})\quad,
\end{eqnarray}
where even order derivatives of the functions $f(x)$ and $s(x)$ have
disappeared. This is the primary reason for defining the averages in
Eqs.~(\ref{eq:thetabar}) and (\ref{eq:mubar}) in this particular way. Defining
the parameters
\begin{equation}
f_{m,p}=f_{m,p}(\overline{\mu},\overline{\theta}) = (-1)^m\int d\overline{x}~
    \overline{x}^p f^{(m)}(\overline{x})
    T(\overline{\mu}+\overline{x}k_B\overline{\theta})\quad,
\label{eq:fdefined}
\end{equation}
which are only functions of the contact parameters $\overline{\mu}$ and
$\overline{\theta}$, the currents can be expressed as
\begin{eqnarray}
I &=& 2\frac{(-e)}{h} k_B\overline{\theta} \sum_{m=0}^\infty \sum_{p=0}^{2m+1}
       \frac{f_{2m+1,p} \epsilon^p \delta^{2m+1-p}}{2^{2m} p!(2m+1-p)!} \quad,
    \label{eq:chexp} \\
I_S &=& 2\frac{k_B}{h} k_B\overline{\theta} \sum_{m=0}^\infty \sum_{p=0}^{2m+1}
       \frac{[f_{2m+1,p+1}-2m f_{2m,p}]\epsilon^p \delta^{2m+1-p}}{2^{2m} p!(2m+1-p)!}
       \quad.
    \label{eq:enexp}
\end{eqnarray}
This is the desired expansion of currents in terms of the driving forces
$\epsilon$ and $\delta$ with the coefficients being functions of the average
quantities $\overline{\mu}$ and $\overline{\theta}$.

One notable property of the equations (\ref{eq:chexp}) and (\ref{eq:enexp}) is
that only the odd powers of the driving forces combined together appear in
those expressions. This implies that if both driving forces change sign
$\epsilon \rightarrow -\epsilon$ and $\delta \rightarrow -\delta$ then the
charge and entropy currents change direction. Including only up to the third
order terms in the expansions we have
\begin{eqnarray}
  \nonumber I&=&2\frac{(-e)}{h}k_B\overline{\theta}\left(f_{10}\delta+f_{11}\epsilon \right.\\
    & & \left. +\frac{1}{24}\left(f_{30}\delta^3+3f_{31}\delta^2\epsilon+3f_{32}\delta\epsilon^2
      +f_{33}\epsilon^3 \right)+\cdots \right) \\
  I_S&=&2\frac{k_B^2\overline{\theta}}{h}\left(f_{11}\delta+f_{12}\epsilon+
    \frac{1}{24}\left( (f_{31}-2f_{20})\delta^3+3(f_{32}-2f_{21})\delta^2\epsilon \right.\right. \nonumber\\
   & & \qquad\left.\left.
      +3(f_{33}-2f_{22})\delta\epsilon^2 + (f_{34}-2f_{23})\epsilon^3 \right)
      +\cdots \right)
\end{eqnarray}
These equations give the currents for arbitrary values of the temperature and
potential differences. However, measurements are rarely carried out for
arbitrary $\Delta\theta$ and $V$. Electrical conductance and Peltier effect
measurements are carried out at isothermal conditions while the thermal
conductance and Seebeck effect measurements are done with zero electrical
current. But, the equations above is a starting point for each particular
phenomenon. In the following, only the Seebeck and Peltier effects are
investigated.

\section{Seebeck Effect}

In the Seebeck effect, a temperature difference creates a potential difference
across the point contact when there is no electrical current ($I=0$). This
potential difference can be expressed in dimensionless form as
\begin{equation}
  -\delta=\sigma_1 \epsilon +\sigma_3 \epsilon^3 +\sigma_5 \epsilon^5+\cdots
\end{equation}
where the first two coefficients are
\begin{eqnarray}
  \sigma_1 &=& \frac{f_{11}}{f_{10}}  \\
  \sigma_3 &=& \frac{1}{24f_{10}}\left(f_{33}-3f_{32}\sigma_1+3f_{31}\sigma_1^2-f_{30}\sigma_1^3\right)
  \label{eq:sigmathreedef}
\end{eqnarray}
In terms of $V$ and $\Delta\theta$ the series expansion is
\begin{equation}
  -V = S_1 \Delta\theta+S_3\Delta\theta^3+S_5\Delta\theta^5+\cdots
  \label{eq:seebexp}
\end{equation}
where
$$ S_{m} = \frac{k_B}{(-e)}\frac{1}{\theta_A^{m-1}}\sigma_{m} \qquad m=1,3,5,\ldots $$
Appearance of only the third order terms in Eq.~(\ref{eq:seebexp}) implies that
when the temperatures of the two reservoirs are exchanged (in other words the
sign of $\Delta\theta$ is changed without changing $\theta_A$ and
$\overline{\theta}$), the induced potential difference due to the Seebeck
effect is reversed.

The nonlinear terms in Eq.~(\ref{eq:seebexp}) becomes significant when
$$ \Delta\theta_\mathrm{threshold}\sim \sqrt{\left\vert\frac{S_1}{S_3}\right\vert}\quad.$$
It is possible to get a theoretical estimate of this quantity in the small
temperature limit, when $k_B\overline{\theta}\ll E_L$, where $E_L$ is the
energy range where $T(E)$ changes by one. In this case, the Taylor series
expansion
$$T(\overline{\mu}+xk_B\overline{\theta})\approx T(\overline{\mu})+xk_B\overline{\theta} T^\prime(\overline{\mu})$$
in Eq.~(\ref{eq:fdefined}) gives the following approximate expressions for
$\sigma_1$ and $\sigma_3$
$$ \sigma_1\approx \frac{\pi^2}{3} \frac{T^\prime}{T} k_B\overline{\theta}\quad,\qquad
\sigma_3\approx \frac{\pi^2}{12}\frac{T^\prime}{T} k_B\overline{\theta}\quad.
$$
The threshold level for nonlinearity is then
$$ \Delta\theta_\mathrm{threshold}\sim 2\theta_A=\theta_L+\theta_R\quad.$$
Since $\Delta\theta$ can never go above this level, the nonlinearities in the
Seebeck effect are always small.\cite{dzurak} For this reason, the expansion
(\ref{eq:seebexp}) is appropriate for almost all nonlinear cases. For the
opposite, high temperature limit, numerical calculations of the Seebeck
coefficients indicates that the threshold expression given above does not
change much.

\begin{figure}[htb]
\includegraphics [scale=0.45]{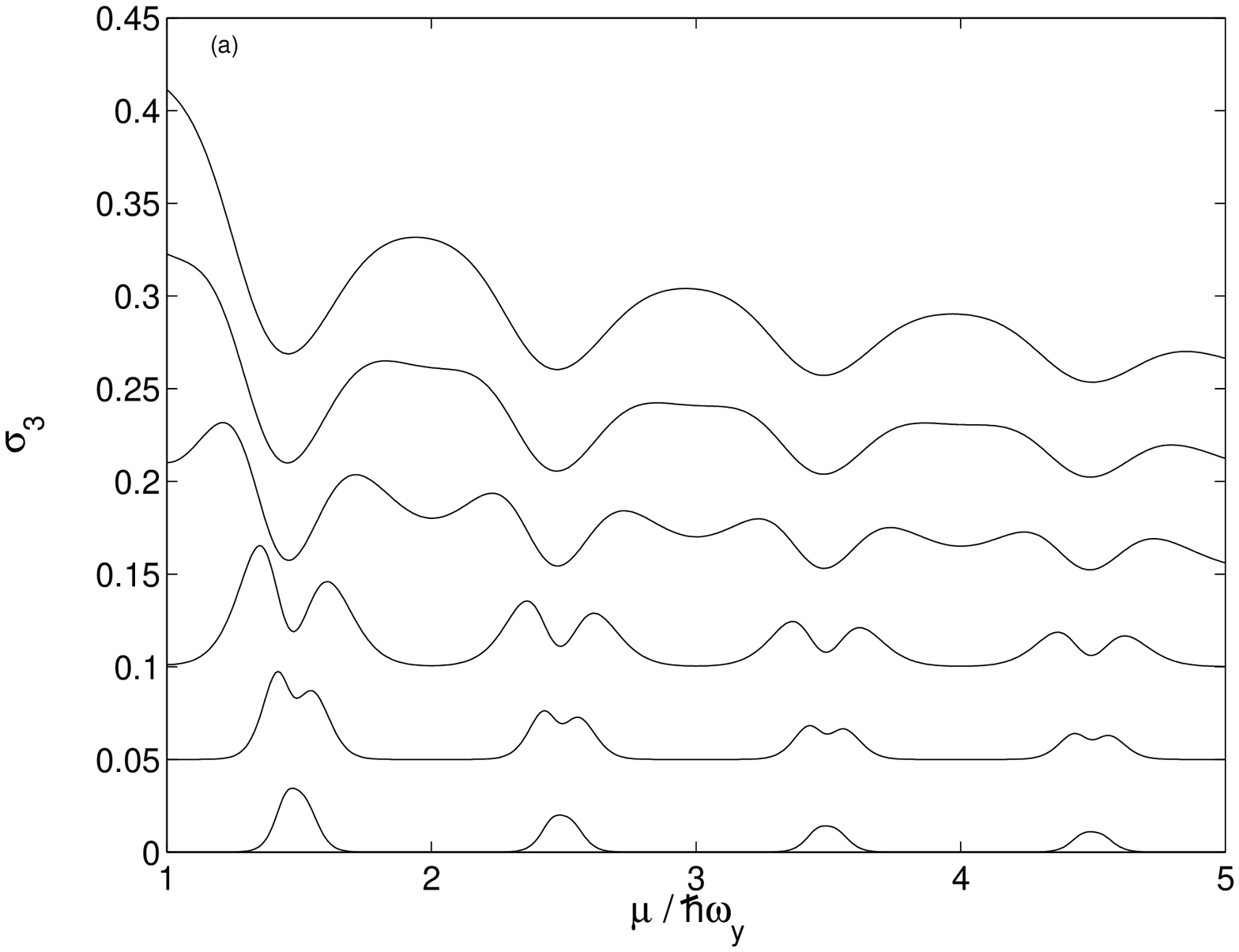}\\
\includegraphics [scale=0.45]{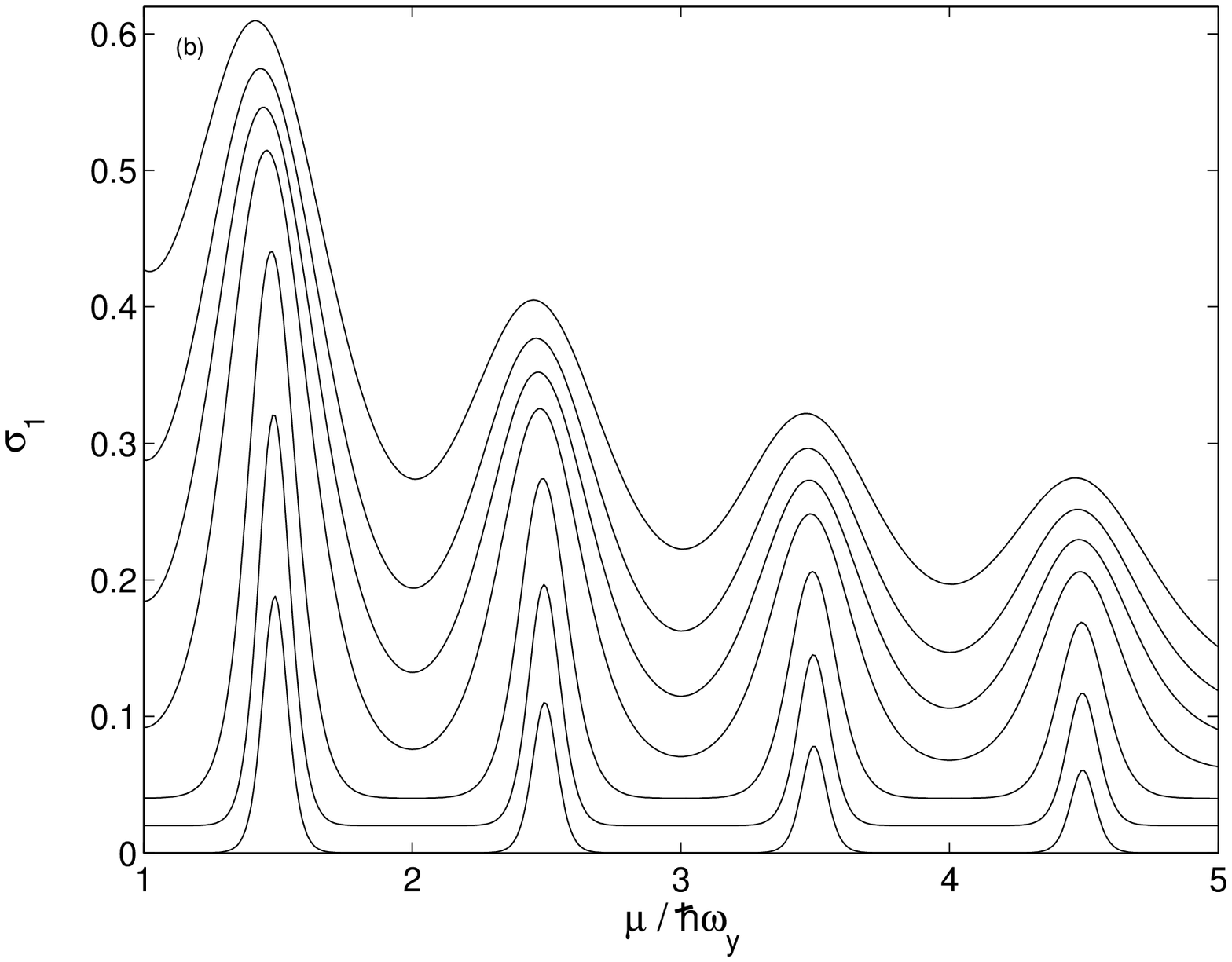}
\caption{\label{fig:seebthree1} (a) The third order Seebeck
coefficient, $\sigma_3=(-e)\theta_A^2/k_B~S_3$, is plotted as a
function of average chemical potential $\overline{\mu}$ for
$\omega_y/\omega_x=6$ and $k_B\overline{\theta}/\hbar\omega_y=$ 0.01, 0.02, 0.04,
0.08, 0.105 and 0.125 (from bottom to top). Each curve is shifted
by 0.05 units for clarity. (b)For comparison the linear Seebeck
coefficient, $\sigma_1=(-e)/k_B~S_1$, is plotted for the same set
of parameters. Each curve is shifted by 0.02 units and the
temperature increases from bottom to top. }
\end{figure}

\begin{figure}
\includegraphics[scale=0.45]{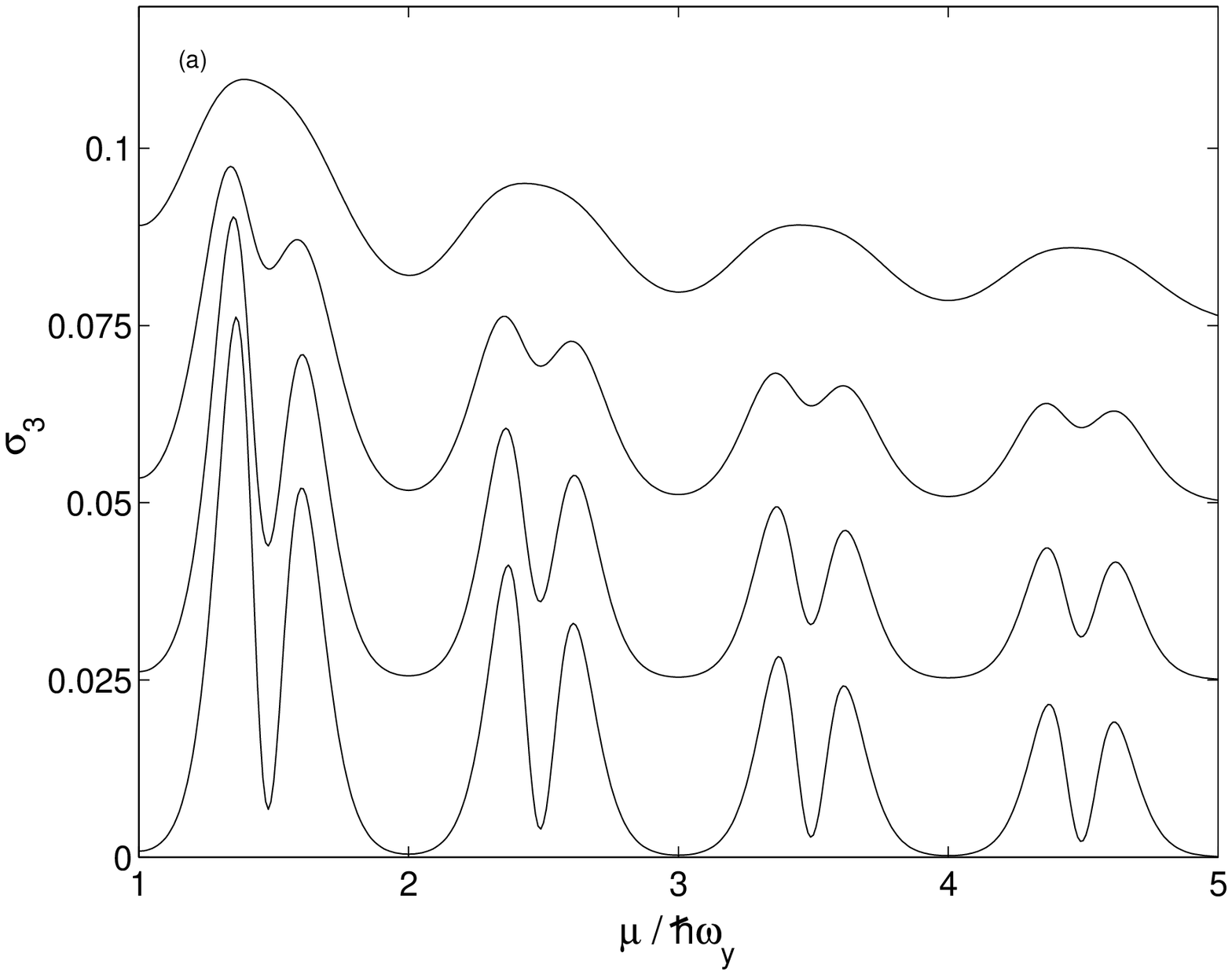}\\
\includegraphics[scale=0.45]{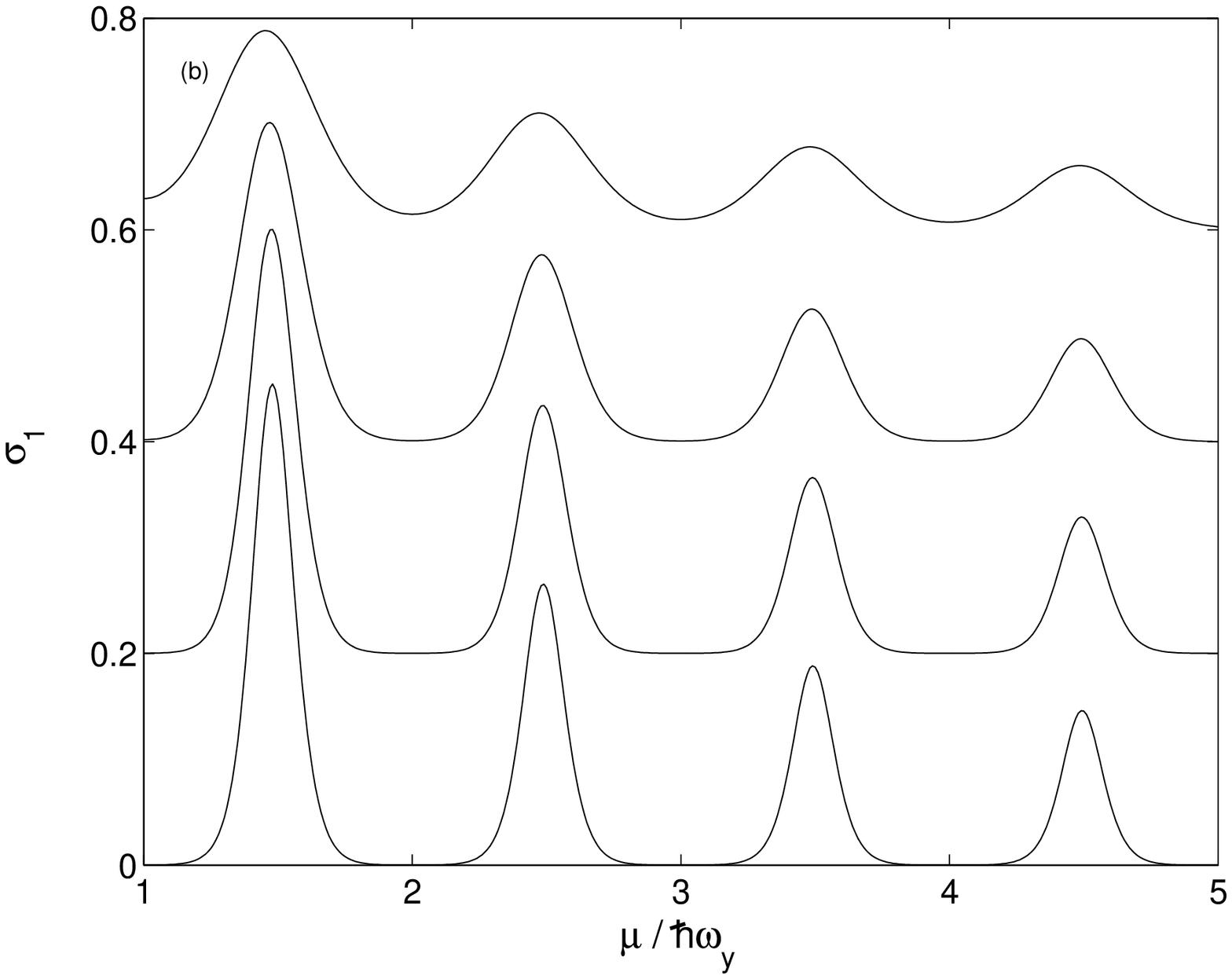}
\caption{\label{fig:seebthree2} (a) The third order Seebeck coefficient,
$\sigma_3=(-e)\theta_A^2/k_B~S_3$, is plotted as a function of average chemical
potential $\overline{\mu}$ for $k_B\overline{\theta}/\hbar\omega_y=0.04$ and
$\omega_y/\omega_x=$ 1.5, 3, 6 and 12 (from top to bottom) respectively. Each
plot is shifted by 0.025 units for clarity. (b)For comparison, the linear
Seebeck coefficient, $\sigma_1=(-e)/k_B~S_1$, is plotted for the same set of
parameters. Each plot is shifted by 0.2 units and $\omega_y/\omega_x$ ratio
increases from top to bottom. }
\end{figure}

As for the general behavior of $S_3$, we calculate it for a contact defined by
the saddle potential
$$ V(x,y) = -\frac{1}{2} m \omega_x^2 x^2 + \frac{1}{2} m \omega_y^2 y^2\quad. $$
For this case the energy dependent transmission probability for the $n$th
transverse mode ($n=0,1,2,\ldots$) is
$$ T_n(E) =\frac{1}{1+\exp\left(-\frac{2\pi}{\hbar\omega_x}[E-\hbar\omega_y(n+\frac{1}{2})]\right)}\quad.$$
In Fig.~\ref{fig:seebthree1}, $S_3$ is plotted against $\overline{\mu}$ for
this potential. At sufficiently low temperatures, third order Seebeck
coefficient, $S_3$, has single peaks coincident with the peaks of $S_1$. When
the temperature is increased, these peaks start to split into two. This change
happens around $k_B\overline{\theta}/\hbar\omega_x \sim 0.08$. It is observed that the distance
between the peaks is proportional to the temperature. For this reason, with
increasing temperature, the structure develops into two separate peaks. Also,
the widths of the peaks increase proportionally with the temperature.
Inevitably, when the temperature is increased further (around
$k_B\overline{\theta}/\hbar\omega_y \sim 0.08$), each peak of the pair starts overlapping with
the peaks of the neighboring steps. For this reason, in this high temperature
regime the nonlinearity in the Seebeck effect becomes more significant away
from the steps (at the plateaus of the electrical conductance). Same graphs are
shown in Fig.~\ref{fig:seebthree2} for different values of $\omega_y/\omega_x$
ratio. It can be seen that $S_3$ has single peaks for small values of
$\omega_y/\omega_x$ ratio (around $\omega_y/\omega_x\sim 1$), and peak
splitting occurs for larger values of the $\omega_y/\omega_x$ ratio.

In all cases it can be seen that $S_3$ is always negative ($\sigma_3$ is always
positive) and never changes sign. It implies that the nonlinearity increases
the generated Seebeck voltage further than the linear term alone suggests. Note
that this feature of $S_3$ is not apparent from its definition,
Eqn.~(\ref{eq:sigmathreedef}). This appears to be a model dependent feature.
Especially if $T(E)$ may decrease for some energies, $S_3$ may display sign
changes. But for the saddle potential model and for all parameter ranges
investigated in this study, $S_3$ is found to have the same sign.

\section{Peltier Effect}

The Peltier heat is defined as the heat carried $\dot{Q}=\theta I_S$ by the
charge current $I$ at isothermal conditions ($\theta_L=\theta_R=\theta$). The
expansion of the Peltier heat and the charge current in terms of the $\delta$
parameter is
\begin{eqnarray}
  \dot{Q} &=& 2 \frac{(k_B\theta)^2}{h} \left(f_{11}\delta
    +\frac{1}{24}(f_{31}-2f_{20})\delta^3
    +\frac{1}{1920}(f_{51}-4f_{40})\delta^5+\cdots\right)\quad,
    \label{eq:expan}  \\
  I &=& 2 \frac{(-e)}{h} \left(f_{10}\delta+\frac{1}{24}f_{30}\delta^3
    +\frac{1}{1920}f_{50}\delta^5+\cdots\right)\quad.
\end{eqnarray}
Both of these expressions can be used to expand $\dot{Q}$ as a power series in
the current $I$
\begin{equation}
\dot{Q} = \Pi_1 I +\Pi_3 I^3 + \Pi_5 I^5 +\cdots\quad, \label{eq:heatexp}
\end{equation}
where the first two terms of the expansion are
\begin{eqnarray}
  \Pi_1 &=& \frac{k_B\theta}{(-e)} \frac{f_{11}}{f_{10}} \quad,\\
  \Pi_3 &=& \frac{h^2}{(-e)^3 k_B\theta} \frac{f_{10}(f_{31}-2f_{20})
    -f_{11}f_{30}}{96 f_{10}^4} \quad,
\end{eqnarray}
The appearance of only the odd powers of the current in the expansion of
$\dot{Q}$ signifies the reversible character of the Peltier heat. The
coefficient $\Pi_1$ is for the linear Peltier effect, which is related to $S_1$
through the Thomson-Onsager relation by $\Pi_1=\theta S_1$.

\begin{figure}
\includegraphics[scale=0.45]{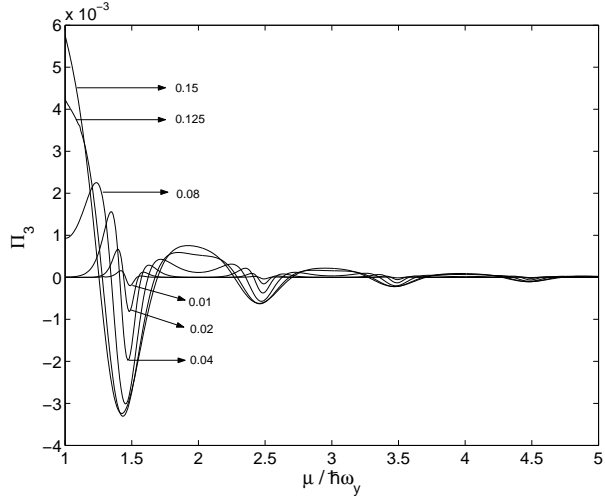}
\caption{\label{fig:peltthree1} The third-order Peltier coefficient $\Pi_3$ (in
arbitrary units) is plotted as a function of average chemical potential
$\overline{\mu}$ for $\omega_y/\omega_x=6$ and different values of temperatures
($k_B\theta/\hbar\omega_y$ values are indicated in the figure). }
\end{figure}

\begin{figure}
\includegraphics[scale=0.45]{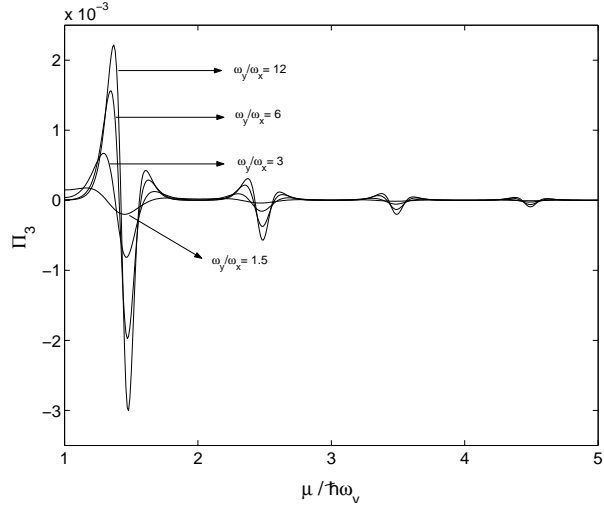}
\caption{\label{fig:peltthree2} The third order Peltier coefficient $\Pi_3$ is
plotted as a function of average chemical potential $\overline{\mu}$ for
$k_B\theta/\hbar\omega_y=0.04$ and different values of $\omega_y/\omega_x$ whose
values are indicated in the figure. }
\end{figure}

The plots of $\Pi_3$ are shown in Fig.~\ref{fig:peltthree1} and
\ref{fig:peltthree2} for the saddle potential model as a function of
$\overline{\mu}$ for different values of parameters $k_B\theta/\hbar\omega_y$ and
$\omega_y/\omega_x$, respectively. For low temperatures, $\Pi_3$ is non-zero
only around the steps of the conductance. But, in contrast to $S_3$, it
displays a change of sign for all parameter values. In particular $\Pi_3$ has
opposite sign at the peaks of $\Pi_1=\theta S$. This behavior is an indication
of the peak splitting\cite{bogachek1,bogachek2} behavior of the Peltier
coefficient under nonlinear currents. In other words, with nonlinear currents,
the Peltier heat decreases at the peaks of the linear Peltier coefficient, but
increases at the foothills of these peaks. Similar to $S_3$, $\Pi_3$ is
extremely small at the plateaus of the conductance for small temperatures, but
when the temperature is higher (comparable to $\hbar\omega_y$) it also becomes
significant at the plateau region. Finally, $\Pi_3$ is significant only around
the first few steps. At higher steps, it is observed that the peak heights are
inversely proportional to the cube of $T(\overline{\mu})$.

To estimate the threshold level for nonlinearity, we use the following
approximations valid in small temperature limit
\begin{eqnarray}
f_{31}-2f_{20} &=& \frac {\pi^2}{3} (k_B\overline{\theta})^3
T^{\prime\prime\prime}\quad, \nonumber\\
f_{11} &=& \frac{\pi^2}{3} (k_B\overline\theta)T' \quad, \nonumber
\end{eqnarray}
in Eq.~(\ref{eq:expan}). Therefore, the nonlinearity sets in when the driving
potential difference is of the order of $eV_{\textrm{threshold}} \sim E_L$.
Since it is possible that the driving potential difference on the contact can
easily exceed this threshold level, in these highly nonlinear cases it will not
be reasonable to use only a few terms of the expansion in
Eq.~(\ref{eq:heatexp}). However, for weakly nonlinear cases, the expansion
above might be useful.

\subsection*{High-order nonlinearity in Peltier effect at small temperatures}
As it was discussed above, highly nonlinear cases cannot be treated
appropriately by the power series expansion discussed here. For this case, we
need to have a better method for evaluating the heat and charge currents
passing through the contact. We consider only the isothermal case appropriate
for the Peltier effect. The charge and entropy currents for this case can be
expressed as
\begin{eqnarray}
I &=& 2\frac{(-e)}{h}\int_{-\infty}^\infty dx (-f^\prime(x))
    [A(\mu_L+xk_B\theta)-A(\mu_R+xk_B\theta)]\quad, \\
I_S &=& 2\frac{k_B}{h}\int_{-\infty}^\infty dx (-xf^\prime(x))
    [A(\mu_L+xk_B\theta)-A(\mu_R+xk_B\theta)]\quad,
\end{eqnarray}
where $A(E)$ is the energy integral of $T(E)$,
$$ A(E)= \int_{-\infty}^E T(E)dE \quad. $$
Assuming small temperatures ($k_B\theta \ll E_L$), the integrands can be
expanded as
$$ A(\mu+xk_B\theta)\approx A(\mu)+xk_B\theta T(\mu)\quad.$$
Keeping only the lowest order terms the currents can be expressed as
\begin{eqnarray}
I &=& 2\frac{(-e)}{h} (A(\mu_L)-A(\mu_R)) \quad,\\
\dot{Q} &=& \frac{2\pi^2}{3h} (k_B\theta)^2 (T(\mu_L)-T(\mu_R)) \quad.
\end{eqnarray}
As was discussed by Bogachek \textit{et al.},\cite{bogachek1,bogachek2} the
differential Peltier coefficient can be expressed as (assuming constant
$\overline{\mu}$)
$$ \Pi_d=\left(\frac{\partial \dot{Q}}{\partial I}\right)_{\overline{\mu}}
= \frac{\pi^2(k_B\theta)^2}{3(-e)}
\frac{T^\prime(\mu_L)+T^\prime(\mu_R)}{T(\mu_L)+T(\mu_R)}\quad.
$$
The peak splitting effect of the nonlinearity can be seen from this expression.
When the potential difference across the contact is less than $E_L$, the
individual peaks of $T^\prime(\mu_L)$ and $T^\prime(\mu_R)$ will join in a
single peak observed in the linear Peltier effect. However, if the potential
difference is more than $E_L$, the contribution of these two terms can be
distinguished since they will form two separate peaks. The distance between the
peaks, then, will be proportional to the applied potential difference.

\section{Conclusions}

The expansions of the charge and entropy currents as a power series in
temperature and potential differences are obtained, assuming that the
transmission probabilities are unchanged by the nonlinearities. The main
advantage of this particular expansion is, through a different definition of
average chemical potential, $\overline{\mu}$, and temperature,
$\overline{\theta}$, some particular terms disappear from the expressions. The
Seebeck and Peltier effects are investigated as special cases and it is found
that the lowest order nonlinearities are of third order in both cases.

In the case of the Seebeck effect, $S_3$ is found to have the same sign as
$S_1$. Although at low temperatures $S_3$ is found to be simply proportional to
$S_1$, its peaks split into two at high temperatures. If $k_B\overline{\theta}$
is comparable to the energy difference between the successive sub-bands, these
peaks may join with the peaks of the neighboring steps, creating an unusual
appearance where $S_3$ has maxima at the plateaus of the conductance and minima
at the steps. In all cases, it is found that the nonlinear signal is small
compared to the linear one.

For the case of the Peltier effect, $\Pi_3$ changes sign as the gate voltage is
changed for all parameter values. The main shortcoming of the expansion
developed here is that in this case the potential difference driving the
current may be chosen above the threshold level for nonlinearity. In such a
case, the expansion is useless as more and more terms have to be added up to
obtain the correct response. In the small temperature limit, an alternative
expression has been developed for the differential Peltier coefficient that is
also valid for highly nonlinear cases.


\end{document}